\documentclass[12pt]{article}
\pdfoutput=1

\usepackage{draft} 
\usepackage{hyperref}
\usepackage{graphicx,color,subfig}
\usepackage{cite}
\usepackage{mciteplus}
\usepackage{skak}
\usepackage{empheq}
\usepackage{anyfontsize}
\usepackage{amsthm}
\newtheorem{theo}{Theorem}
\newtheorem{define}{Definition}

\newtheorem{lemma}{Lemma}
\newtheorem{corollary}{Corollary}
\usepackage{comment}

\DeclareFontFamily{OT1}{pzc}{}
\DeclareFontShape{OT1}{pzc}{m}{it}{<-> s * [1.10] pzcmi7t}{}
\DeclareMathAlphabet{\mathpzc}{OT1}{pzc}{m}{it}

\def\be#1\ee{\begin{align}#1\end{align}}

\begin{document}

\unitlength = .8mm

\begin{titlepage}

\begin{center}

\hfill \\
\hfill \\
\vskip 1cm

\title{Bootstrap, Markov Chain Monte Carlo, and LP/SDP Hierarchy for the Lattice Ising Model}

\author{Minjae Cho${}^\#$, Xin Sun${}^*$}

\address{
${}^\#$Princeton Center for Theoretical Science, Princeton University, \\ Princeton, NJ 08544, USA
\\
${}^*$Beijing International Center for Mathematical Research, Peking University
}

\email{${}^\#$minjae@princeton.edu, ${}^*$xinsun@bicmr.pku.edu.cn}

\end{center}

\abstract{Bootstrap is an idea that imposing consistency conditions on a physical system may lead to rigorous and nontrivial statements about its physical observables. In this work, we discuss the bootstrap problem for the invariant measure of the stochastic Ising model defined as a Markov chain where probability bounds and invariance equations are imposed. It is described by a linear programming (LP) hierarchy whose asymptotic convergence is shown by explicitly constructing the invariant measure from the convergent sequence of moments. We also discuss the relation between the LP hierarchy for the invariant measure and a recently introduced semidefinite programming (SDP) hierarchy for the Gibbs measure of the statistical Ising model based on reflection positivity and spin-flip equations.
}

\vfill
\par\noindent\rule{180pt}{0.4pt}\\
\noindent{\footnotesize ${}^*$On leave from the University of Pennsylvania}
\end{titlepage}

\eject

\begingroup
\hypersetup{linkcolor=black}

\tableofcontents

\endgroup

\section{Introduction}
Statistical Ising model is defined by a specific probability measure, called the Gibbs measure, over the space of spin configurations on a lattice.\footnote{See e.g. \cite{glimmjaffe,liggett2004interacting} for an introduction to the subject.} Despite the simplicity of its definition, it exhibits surprisingly rich dynamics which has driven developments of several important branches of math and physics. In particular, the existence of the phase transition in two and three dimensions provides an outstanding example of dramatic physical phenomena that may take place in the infinite volume systems.

Even though there exist analytic solutions in some special cases \cite{PhysRev.65.117,PhysRevB.13.316}, the statistical Ising model under general temperature and external magnetic field in two and higher dimensions still remains unsolved - for example, the value of the critical temperature in three dimensions is unknown. Traditionally, numerical estimates of various quantities were obtained using the Monte Carlo simulations, where probable spin configurations (on a finite lattice though) are sampled over based on the Gibbs measure.\footnote{See e.g. \cite{MCstat} for an introduction to the Monte Carlo simulations of statistical physics.} Markov Chain Monte Carlo (MCMC) is one of the standard dynamical procedures defining such sampling, which is also called the stochastic Ising model when restricted to the Ising model.

Alternatively, there is another approach called "bootstrap" where consistency conditions of the model are imposed and the corresponding consequences are studied. In particular, the conformal bootstrap program has been very successful in studying the continuum theory that arises at the criticality in two and three dimensions. Unitarity, conformal symmetry, and the consistency conditions known as crossing equations provide exact solutions in two dimensions \cite{BELAVIN1984333}, and rigorous and highly tight bounds on physical data in three dimensions with the help of semidefinite programming (SDP) \cite{Rattazzi:2008pe,El-Showk:2012cjh,Kos:2016ysd,Poland:2018epd}. Recently in \cite{Cho:2022lcj}, a different bootstrap approach (labeled $BS_2'$ in this work) was applied directly to the statistical Ising model on the infinite lattice, where reflection positivity and spin-flip equations satisfied by the Gibbs measure were represented as a SDP problem and provided rigorous (and sometimes highly tight) bounds on the spin correlators. It is worth mentioning that the very definition of the Gibbs measure on the infinite lattice given by the DLR equations \cite{Dobruschin1968TheDO,1969CMaPh..13..194L} allows for such a bootstrap formulation very naturally.

An obvious but essential fact about MCMC is that, by construction, the Gibbs measure is guaranteed to be an invariant measure of MCMC. Moreover, under the assumption of translation invariance, every invariant measure is also Gibbs (see Theorem \ref{invGibbs}). Therefore, it is natural to pose another bootstrap problem (labeled $BS_1$ in this work) where probability bounds (stating that the measure is a probability measure) and the invariance condition (stating that the measure is invariant under the Markov chain dynamics) are imposed as bootstrap conditions.\footnote{See e.g. \cite{hernández2012markov} for an introduction to the bootstrap approach for the invariant measures of Markov chains. Also see \cite{2018arXiv180708956K,2015arXiv151205599F,2018PhLA..382..382T} for more recent works on bootstrapping (stochastic) dynamical systems. We thank Hamza Fawzi for pointing out relevant works to us.} These conditions must be met by the Gibbs measure and thus should be compatible with the bootstrap problem based on reflection positivity and spin-flip equations in the sense that they should share common solutions. As we discuss in section \ref{sec:bootstrap}, $BS_1$ is described by a linear programming hierarchy while $BS_2'$ is described by a SDP hierarchy, where the bootstrap conditions at the lower level are part of those at higher levels in both cases. For any choice of the transition rate for MCMC, the set of invariance equations in $BS_1$ will manifestly be a proper subset of the set of spin-flip equations in $BS_2'$ at each level in the two hierarchies.

The hierarchy of LP/SDP encountered in this work is a special case of the Lasserre hierarchy which is much studied in the optimization literature.\footnote{See \cite{10.1007/3-540-45535-3_23,doi:10.1137/S1052623400366802} for the original works by Lasserre and \cite{Laurent2009} for a comprehensive survey.} Statistical mechanical systems provide a unique setup for the Lasserre hierarchy where the number of the polynomial variables is infinite as opposed to finite. One immediate question is the convergence of such hierarchy and it was conjectured in \cite{Cho:2022lcj} that the lower and upper bounds on spin correlators obtained from $BS_2'$ converge to each other as the hierarchy level increases, when there is a single phase. As the main result of this work, we will show the asymptotic convergence of the LP hierarchy of $BS_1$ in the sense that the solutions to the LPs converge to moments of an invariant probability measure of MCMC. As an intermediate step, we will also discuss the relevant moment problem over the space of spin configurations on the infinite lattice. Similar convergence statement for $BS_2'$ remains unclear to us at the moment. Instead, we will define the bootstrap problem $BS_2$ by equipping $BS_2'$ with probability bounds of $BS_1$, which in practice requires only little extra computational cost while the convergence still holds true.

This paper is organized as follows. We first review the definitions and relevant theorems of the stochastic and statistical Ising model in section \ref{sec:review}. They will naturally lead to the bootstrap problems $BS_1$ and $BS_2$ which we introduce in section \ref{sec:bootstrap}. In section \ref{sec:convergence}, we discuss the moment problem for the spin configurations on the infinite lattice and show the convergence of $BS_1$. We provide the bounds obtained by different bootstrap approaches in section \ref{sec:practice} and end with further discussions in section \ref{sec:diss}.

\section{Review of the statistical and stochastic Ising model}\label{sec:review}
In this section, we will review the definitions of the statistical and stochastic Ising model and their relations, and rephrase their properties in terms of the polynomial moments. We will mostly follow \cite{liggett2004interacting} where the details of the theorems and proofs may be found. Even though this section collects very elementary facts about the statistical and stochastic Ising model, showing that they are all satisfied by the solution of the bootstrap problem to be defined later will be the main result of this work, which provides several interesting implications.

\subsection{Probability space for the Ising model}
In this work, we are going to work on the infinite $d$-dimensional hypercubic lattice $\Lambda={\mathbb Z}^d$. At each lattice site $i\in\Lambda$, we have a spin degree of freedom $s_i\in\{-1,1\}$. The state space $S$ is the set of all possible spin configurations $s$ over the lattice $\Lambda$: $S=\{-1,1\}^\Lambda$. The space $S$ is compact and metrizable, with the metric $M:S\times S\rightarrow\mathbb{R}$ given by $M(s,s')=\sum_{i\in\Lambda}2^{-||i||_\infty}\left(1-s_is'_i\over2\right)$ for $s,s'\in S$, where $||\cdot||_\infty$ is the $L_\infty$-norm \cite{liggett2004interacting,friedli_velenik_2017}. For example, as can be easily seen, $M(s,s)=0$ and $M(s,s')<\infty$ for all $s,s'\in S$. The topology of the space $S$ and notions such as continuous functions on $S$ follow from the explicit form of the metric $M$. We will be interested in a specific set of probability measures on the sample space $S$. In order to define the event space, we first define the following.

\begin{define}\label{def:eventspace}
Let $A$ be a finite subset of $\Lambda$, and $u_i\in\{-1,1\}$ for $i\in A$ be a specific spin assignments over the lattice sites of $A$. An event $E(\{u_i\}_{i\in A})$ is defined as the following set of spin configurations:
\ie\label{eventDef}
E(\{u_i\}_{i\in A})=\{s\in S~|~s_i=u_i,~\forall i\in A \}.
\fe
\end{define}

In other words, $E(\{u_i\}_{i\in A})$ is the set of all spin configurations whose spins at lattice sites of $A$ agree with $u_i$. Note that the above definition applies to the case $A=\varnothing$: $E\left(\{u_i\}_{i\in\varnothing}\right)=S$. The event space is going to be the union of the events for all finite subsets $A\in\Lambda$ and all possible spin assignments $u_i$ over them, together with the empty set.

\begin{define}
The event space $V$ is the $\sigma$-algebra generated by the events $E(\{u_i\}_{i\in A})$ for all finite subsets $A\in\Lambda$ and all possible spin assignments $\{u_i\}_{i\in A}$ over them.
\end{define}

A probability measure over $S$ and $V$ is defined as follows.

\begin{define}
For the sample space $S$ and the event space $V$ defined as above, a probability measure over them is a function $\R:V\rightarrow[0,1]$ such that
\\
\textbullet ~ $\R\left(\varnothing\right)=0$ and $\R\left(S\right)=1$,
\\
\textbullet ~ given any countable collection of pairwise disjoint events $\{E_a\}_{a=1}^\infty\subseteq V$, $\R$ is countably additive: $\R\left(\bigcup\limits_{a=1}^\infty E_a\right)=\sum\limits_{a=1}^\infty\R(E_a)$.

\end{define}

$\R\left(E_a\right)$ has the interpretation of probability that the event $E_a$ happens. Later when we try to construct a probability measure for the statistical and stocahstic Ising model from the candidate moments obtained by LPs, it will be important to check that all the requirements in the above definition are satisfied.

In order to define the expectation values associated with a probability measure $\R$, we introduce the indicator functions.

\begin{define}
Given an event $E(\{u_i\}_{i\in A})$, the corresponding indicator function $F(\{u_i\}_{i\in A},\cdot):S\rightarrow\{0,1\}$ is given by
\ie\label{indicator}
F(\{u_i\}_{i\in A},s)=\prod_{i\in A}\left({1+u_is_i\over2}\right).
\fe

\end{define}

As the name suggests, the indicator function for the event $E(\{u_i\}_{i\in A})$ evaluated on a spin configuration $s\in S$ is equal to 1 if the spin assignments of $s$ agrees with $\{u_i\}_{i\in A}$ over $A$, and 0 otherwise. The construction of the expectation value then proceeds as usual.

\begin{define}
Given a probability measure $\R$ over the sample space $S$ and the event space $V$, and a function $f:S\rightarrow{\mathbb R}$, the expectation value of $f$ given by $\R$ is
\ie
\langle f(s)\rangle = \int_Sf(s)d\R.
\fe

\end{define}

\subsection{The statistical and stochastic Ising model}\label{sec:isingReview}
We are now ready to define the statistical and stochastic Ising model. For any given site $i\in\Lambda$, its nearest neighbors are the collection of sites $n(i):=\{j\in\Lambda~|~||i-j||_1=1\}$, where $||\cdot||_1$ is the $L_1$-norm. The Ising model is local in the sense that its probability measure is defined using only the nearest neighboring spins.

\begin{define}\label{gibbsDef}
The Gibbs measure $g$ of the statistical Ising model on the lattice $\Lambda={\mathbb Z}^d$ at couplings $J\in\mathbb R$ and $h\in\mathbb R$ is a probability measure over the sample space $S$ and the event space $V$ such that:
\\
$~~~$ given any lattice site $i\in\Lambda$, any finite subset $T\subset\Lambda$ such that $n(i)\subset T$ and $i\notin T$, any spin assignments $\{u_k\}_{k\in T}$ over $T$, and any spin assignment $u_i$ at $i\in\Lambda$,
\ie
g\left(E\left(\{u_k\}_{k\in T\cup\{i\}}\right)\right)={g\left(E\left(\{u_k\}_{k\in T}\right)\right)\over1+e^{-2\left(h u_i+J\sum_{j\in n(i)}u_i u_j \right)}}.
\fe
The set of all Gibbs measures at couplings $J$ and $h$ is denoted as $\Gamma_{J,h}$.
\end{define}
This definition is equivalent to the traditional one given by the DLR equations \cite{Dobruschin1968TheDO,1969CMaPh..13..194L}. In case $g\left(E\left(\{u_k\}_{k\in T}\right)\right)\neq0$, this is equivalent to saying that the conditional probability that the spin $s_i$ at $i\in\Lambda$ takes the value $u_i$, given the spin assignments $\{u_k\}_{k\in T}$ over $T$ which in particular includes the nearest neighbors of $i$, is given by $\left(1+e^{-2\left(h u_i+J\sum_{j\in n(i)}u_i u_j \right)}\right)^{-1}$. When $J\geq0$ and $h\geq0$, the statistical Ising model is called ferromagnetic, and we are going to focus only on the ferromagnetic case in this work.

The above definition using the conditional probability agrees with the conventional definition of the statistical Ising model on the finite lattice $\Lambda_f$ (Proposition 1.8 in Chapter IV of \cite{liggett2004interacting}), which is described by the partition function
\ie
Z=\sum_{s\in S}\text{exp}\left(J\sum_{(i,j)}s_is_j+h\sum_i s_i\right),
\fe
and probability measure
\ie\label{gfinite}
g_f\left(E\left(\{u_k\}_{k\in \Lambda_f}\right)\right)={1\over Z}\text{exp}\left(J\sum_{(i,j)}u_iu_j+h\sum_i u_i\right),
\fe
where $\sum_{(i,j)}$ means that the sum is over all the nearest neighbor pairs $(i,j)$.

It is very important that depending on the value of $J$ and $h$ (and also the dimension $d$), there may be more than one Gibbs measure satisfying Definition \ref{gibbsDef}! This is the hallmark of the phase transition which may take place only on the infinite lattice.

Now, we turn to the definition of the stochastic Ising model.

\begin{define}\label{stochDef}
Given the couplings $J\in\mathbb R$ and $h\in\mathbb R$, the stochastic Ising model is a Markov chain on the state space $S$ such that:
\\
\textbullet~on every lattice site of $\Lambda={\mathbb Z}^d$, a Poisson clock is placed, namely each site is associated with a Poisson point process\footnote{A poisson point process is a random collection of points $\{S_1,S_2,\cdots\}$ on $(0,\infty)$ where $\{S_{n+1}-S_n: n\ge 1\}$ are identically independently distributed exponential random variables.} where the occurrence of points is viewed as the times when the clock at that site rings;
\\
\textbullet~if the current state is given by $s\in S$ and the Poisson clock at the site $i\in\Lambda$ rings, the state $s$ makes a transition to another state $s'\in S$ with a strictly positive transition rate $c(i,s)$ where $s'_j=s_j,~\forall j\in \Lambda\setminus\{i\}$, and $s'_i=-s_i$;
\\
\textbullet~the function $c(i,s)e^{h s_i+J\sum_{j\in n(i)}s_i s_j }$ does not depend on the value of $s_i$.

\end{define}

On the finite lattice, the equivalent of placing Poisson clocks is to randomly choose a site with a uniform distribution at each discrete time as in Monte Carlo simulation. On the infinite lattice, we instead place Poisson clocks on every site to "uniformly" choose which spin to update. In particular, the expected number of ringings of a Poisson clock grows linearly in time.

Note that we did not specify the transition rate (or the transition probability) $c(i,s)$. The key idea is that as long as $c(i,s)$ satisfies the last condition in Definition \ref{stochDef}, the objects of interest (which we will introduce soon) will be independent of the specific choice of $c(i,s)$. Independence on the value of $s_i$ is equivalent to saying that the function is even in $s_i$. Popular choices for $c(i,s)$ are $c(i,s)=\text{exp}\left(-h s_i-J\sum_{j\in n(i)}s_i s_j\right)$ and $c(i,s)=\left(1+\text{exp}\left(2h s_i+2J\sum_{j\in n(i)}s_i s_j\right)\right)^{-1}$. Later in section \ref{sec:practice}, we will work with the following choice:
\ie\label{transRate}
c(i,s)=c^*(i,s):=C\left(1+\text{exp}\left(-2h s_i-2J\sum_{j\in n(i)}s_i s_j\right)\right),
\fe
where $C$ is a constant depending on $d,J,$ and $h$ whose details will not matter for us. One possible choice would be $C=1/\left(1+\text{exp}(4dJ+2h)\right)$.

When we apply the above definition to the case where $\Lambda$ is finite, we obtain the traditional Markov chain (sometimes called the Glauber dynamics) which is used to perform the Monte Carlo simulation of the Ising model, known as MCMC. The last condition in Definition \ref{stochDef} is nothing but the detailed balance equation for the probability measure $g_f$ in (\ref{gfinite}). The ergodicity theorem states that $g_f$ is indeed the unique invariant measure of the Markov chain. Of course for our case where $\Lambda$ is infinite, the set of invariant measures needs not be a singleton.

\begin{define}
A probability measure $\R$ over the sample space $S$ and the event space $V$ is an invariant measure of the stochastic Ising model if
\ie
\sum_{i\in S}\bigg\langle c(i,s)\left(f(\overline{s}^i)-f(s)\right)\bigg\rangle=\sum_{i\in S}\int_S c(i,s)\left(f(\overline{s}^i)-f(s)\right)d\R=0,~~\forall f\in D(S),
\fe
where $\overline{s}^i\in S$ is defined by $\left(\overline{s}^i\right)_j=s_j,~\forall j\in\Lambda\setminus\{i\}$, and $\left(\overline{s}^i\right)_i=-s_i$.

We denote by $\Pi_{J,h}$ the set of all invariant measures of the stochastic Ising model at couplings $J$ and $h$.

\end{define}

The definition of the space of functions $D(S)$ (sometimes called the core of the Markov chain) can be found in Chapter I of \cite{liggett2004interacting}. For us, the only relevant facts about $D(S)$ are that it is a dense subset of the set $C(S)$ of continuous functions on $S$, and the set $P(S)$ of polynomials in $\{s_i\}_{i\in\Lambda}$ is a subset of $D(S)$. As the name suggests, the invariant measure remains invariant under the time evolution of the Markov chain.

The stochastic Ising model is defined such that the Gibbs measure of the statistical Ising model is a reversible measure.

\begin{define}
A probability measure $\R$ over the sample space $S$ and the event space $V$ is a reversible measure of the stochastic Ising model if
\ie
\bigg\langle c(i,s)\left(f(\overline{s}^i)-f(s)\right)\bigg\rangle=\int_S c(i,s)\left(f(\overline{s}^i)-f(s)\right)d\R=0,~~\forall i\in\Lambda,~\forall f\in C(S).
\fe

We denote by $\Omega_{J,h}$ the set of all reversible measures of the stochastic Ising model at couplings $J$ and $h$.

\end{define}

\begin{theo}\label{reverGibbs} (Theorem 2.14 in Chapter IV of \cite{liggett2004interacting}) Given $J\in\mathbb R$ and $h\in\mathbb R$, $\Omega_{J,h}=\Gamma_{J,h}$.
\end{theo}

Note that a reversible measure is invariant. Furthermore, Theorem \ref{reverGibbs} says that a reversible measure is a Gibbs measure. This is essentially because the reversibility condition and the conditional probability defining the Gibbs measure are equivalent. Also note that Theorem \ref{reverGibbs} does not rely on the specific choice of the transition rate $c(i,s)$. This implies that the set of reversible measures is independent of the choice of $c(i,s)$ as long as the latter satisfies the definition of the stochastic Ising model. A natural question is whether there are invariant measures which are not reversible. It was shown in \cite{1971CMaPh..23...87H} that there are no such measures under the assumption of translation invariance.

\begin{define}
Let $t_p:\Lambda\rightarrow\Lambda$ for $p=1,2,...,d$ be a translation of the lattice sites by one unit in $p$-th direction: $t_p(i)=i+e_p$ where $e_p$ is the unit vector along the $p$-th direction. A probability measure $\R$ over the sample space $S$ and the event space $V$ is translation invariant if $\R(E(\{u_i\}_{i\in A}))=\R(E(\{v_j\}_{t_p(j)\in A}))$ for all events $E(\{u_i\}_{i\in A})$ and $p$, where $v_{t^{-1}_p(i)}=u_i,~\forall i\in A$.
\end{define}

\begin{theo}\label{invGibbs} Let $\R$ be a translation invariant probability measure over the sample space $S$ and the event space $V$. Then, $\R\in\Pi_{J,h}~\Leftrightarrow~\R\in\Omega_{J,h}~\Leftrightarrow~\R\in\Gamma_{J,h}$.
\end{theo}

In fact, it can be shown that for $d=1$ and $d=2$, invariant measures are reversible even in the absence of the translation invariance assumption (see e.g. Chapter IV.5 of \cite{liggett2004interacting}). However, as far as we are aware, this is not established for $d\geq3$.

\subsection{Moments, positivity, invariance, and reversibility}
Later when we formulate the bootstrap problems for the Ising model, the information about a probability measure will be expressed in terms of moments. Therefore, we describe the properties of a probability measure discussed so far in terms of moments in this subsection.

Say that we are given a candidate set of polynomial moments $\langle p(s)\rangle$, $\forall p(s)\in P(S)$. The question is, how do we make sure that they correspond to the expectation values of some probability measure $\R$ satisfying either invariance or reversibility? In the general case of real-valued polynomial moment problems, this type of question remains unsolved. However, as we will see in this work, this question for the Ising model has a definite answer.

We first address the positivity of the candidate measure. Given a candidate set of polynomial moments $\langle p(s)\rangle$, $\forall p(s)\in P(S)$, we know in particular the moments of all the indicator functions because indicator functions $F(\{u_i\}_{i\in A},s)=\prod_{i\in A}\left({1+u_is_i\over2}\right)$ for events $E(\{u_i\}_{i\in A})$ are polynomials themselves. Then, the candidate probability measure $\R$ realizing the given set of moments should satisfy
\ie
\R(E(\{u_i\}_{i\in A}))=\langle F(\{u_i\}_{i\in A},s)\rangle,
\fe
for all events $E(\{u_i\}_{i\in A})$. This is a natural requirement for the candidate measure $\R$ since the value of the measure evaluated on an event has the interpretation of the probability that the event takes place, which in turn should be equal to the expectation value of the corresponding indicator function. Therefore,
\begin{lemma}
A candidate probability measure $\R$ over the sample space $S$ and the event space $V$ is positive only if its candidate moments satisfy $\langle F(\{u_i\}_{i\in A},s)\rangle\geq0$ for all events $E(\{u_i\}_{i\in A})$.
\end{lemma}
Note that the above Lemma states only a necessary condition for the probability measure. Such a condition can be readily checked for the candidate moments $\langle p(s)\rangle$. In contrast to the general polynomial moment problems where the indicator functions are not polynomials and thus require extra conditions to even discuss their moments, the Ising model (and many other statistical models) is particularly simple since the indicator functions are polynomials. Just checking the positivity of the candidate probability measure evaluated on the generators of the event space $V$ is not enough to guarantee that it is indeed a probability measure, since one also has to make sure that countable additivity can be made sense. We will have further discussions on this in section \ref{subsec:momentProblem}.

Next, we turn to the invariance and reversibility conditions for a candidate measure and candidate moments. Given $s'\in S,~s''\in S$ such that $s'\neq s''$, there exists at least one site $i\in \Lambda$ such that $s'_i\neq s''_i$. The polynomial function $s_i$ then separates two points $s'$ and $s''$. Therefore, the set $P(S)$ of polynomials in $\{s_i\}_{i\in\Lambda}$ is a subalgebra of $C(S)$ which separates points in $S$. As already discussed above Definition \ref{def:eventspace}, the space $S$ is compact under the metric $M$. Then, Stone-Weierestrass theorem implies that $P(S)$ is dense in $C(S)$, and also in $D(S)$.\footnote{A notion of the sequential compactness of $S$ and thus the statement that $P(S)$ is dense in $C(S)$ can be found for example in Chapter 6 of \cite{friedli_velenik_2017}.}

The implication of this fact is that the invariance and reversibility for a measure, which by definition require considering the expectation values of arbitrary functions in $D(S)$ and $C(S)$, can be checked by considering only the polynomial moments.

\begin{lemma}\label{invEqPoly}
A probability measure $\R$ over the sample space $S$ and the event space $V$ is an invariant measure of the stochastic Ising model if and only if its polynomial moments satisfy
\ie\label{invPoly}
\sum_{i\in S}\bigg\langle c(i,s)\left(f(\overline{s}^i)-f(s)\right)\bigg\rangle=0,~~\forall f\in P(S).
\fe
\end{lemma}

\begin{lemma}
A probability measure $\R$ over the sample space $S$ and the event space $V$ is a reversible measure of the stochastic Ising model if and only if its polynomial moments satisfy
\ie\label{revPoly}
\bigg\langle c(i,s)\left(f(\overline{s}^i)-f(s)\right)\bigg\rangle=0,~~\forall i\in\Lambda,~\forall f\in P(S).
\fe
\end{lemma}

It may not be immediately obvious how $c(i,s)\left(f(\overline{s}^i)-f(s)\right)$ may be expressed as a polynomial. This is essentially because the spin variables $s_i$ at each site $i\in S$ can take values only in $\{-1,1\}$ and $c(i,s)$ is a local expression around the site $i$ involving only the nearest neighbors so that any reasonable choice of $c(i,s)$ (such as ones discussed around (\ref{transRate})) can be equivalently written as a polynomial of finite number of spin variables. Therefore, (\ref{invPoly}) and (\ref{revPoly}) are indeed equations for polynomial moments.

\section{Bootstrap problems for the stochastic and statistical Ising model}\label{sec:bootstrap}
In this section, we formulate the bootstrap problems for the stochastic and statistical Ising model. Such a formulation is very natural from the definitions of the stochastic and statistical Ising model for two reasons. The first is that the object of interest is a probability measure, whose positivity is a crucial defining property. The second is that any such measure satisfying a given set of equations (invariance or reversibility) is physical. The combination of positivity and equations provides a bootstrap-friendly setup, and it is thus expected that imposing them over the set of candidate measures would lead to rigorous and nontrivial results about the space of physical measures.

%With any choice of the transition rate for the stochastic Ising model, we will observe that the invariance/reversibility conditions of the bootstrap problem for the stochastic Ising model are contained in spin-flip equations for the statistical Ising model.

We first begin by introducing some notations. We define $D_n=\{i\in\mathbb{Z}^d|~||i||_1\leq n-1 \}$ for $n=1,2,3,...$, where $||\cdot||_1$ is the $L_1$-norm. For example, it is a diamond in $d=2$ and octahedron in $d=3$. The hierarchy of LP/SDP for the bootstrap problems originates in part from the hierarchy of $D_n$.

Given two subsets $A\subset\Lambda$ and $B\subset\Lambda$, we write $A\sim B$ if they can be transformed into each other by a symmetry transformation of the lattice $\Lambda={\mathbb Z}^d$ (which are generated by translations, rotations, and reflections). This defines an equivalence relation on the set of finite subsets of $\Lambda$.

Given any finite subset $A\subset\Lambda$, we define the monomials $\underline{s}_A:=\prod_{i\in A}s_i$, and we also define $\underline{s}_{\varnothing}:=1$. For each $n$, we further define $P_n:=\{\sum_{A\in D_n}t^A\underline{s}_A,~t^A\in{\mathbb R} \}$, the set of polynomials in spin variables restricted to $D_n$. In the hierarchy of LP/SDP, the level $n$ LP/SDP will impose constraints on candidate moments for polynomials in $P_n$. Such candidate moments will be denoted as $m_n:P_n\rightarrow{\mathbb R}$.

\subsection{Bootstrapping the invariant measure of the stochastic Ising model}
We now introduce the hierarchy of LPs which provides a series of rigorous bounds on the objective moment of the invariant measure of the stochastic Ising model.

\begin{define}\label{defBS1}
Given $p\in P_m$ for some $m\in{\mathbb N}$, we define the bootstrap problem $BS_1(p)$ as the following hierarchy of LPs:
\\
$~~~$ For each $n\in{\mathbb N}$ (called the level of the $LP$ hierarchy) such that $n\geq m$, we have the LP problem $LP(p,n)$ of minimizing $m_n(p)$ over the space of candidate moments $m_n:P_n\rightarrow{\mathbb R}$ satisfying the following conditions:
\\
\textbullet~ \textbf{Probability bound.} For all the spin assignments $\{u_i\}_{i\in D_n}$ over $D_n$, $0\leq m_n\left(F\left(\{u_i\}_{i\in D_n},s \right)\right)$ where $F\left(\{u_i\}_{i\in D_n},s \right)$ is the corresponding indicator function.
\\
\textbullet~ \textbf{Linearity.} Given any polynomials $q_1\in P_n$ and $q_2\in P_n$, with $\lambda\in\mathbb R$, their moments satisfy linearity: $m_n(q_1+\lambda q_2)=m_n(q_1)+\lambda m_n(q_2)$.
\\
\textbullet~ \textbf{Unit normalization.} $m_n(1)=1.$
\\
\textbullet~ \textbf{Symmetry.} For any $A\subset D_n$ and $B\subset D_n$ such that $A\sim B$, $m_n(\underline{s}_A)=m_n(\underline{s}_B)$.
\\
\textbullet~ \textbf{Invariance.} For any polynomial $f\in P_{n-1}$, the moments satisfy the invariance with respect to the transition rate $c(i,s)$ of the stochastic Ising model in (\ref{transRate}):
\ie\label{sdpInv}
\sum_{i\in D_{n-1}}m_n\left(d^f_i\right)=0,
\fe
where $d^f_i(s):=c(i,s)\left(f(\overline{s}^i)-f(s) \right)$ is an element of $P_n$ due to $s_j^2=1$, $\forall j\in \Lambda$.
\\
\\
$~~~$ The minimum of $m_n(p)$ obtained by $LP(p,n)$ will be denoted as $\langle p\rangle_n^*$. The corresponding candidate moments $m_n(q)$ for polynomials $q\in P_n$ realizing such a minimum (which may not be unique) will be denoted as $\langle q\rangle_n^*$.

\end{define}

A few comments are in order. Firstly, the invariance condition written above makes sense because $f\in P_{n-1}$, and the transition rate $c(i,s)$ depends only on the nearest neighbors of the site $i\in D_{n-1}$, so that $d^f_i$ indeed is an element of $P_n$ and (\ref{sdpInv}) therefore is a linear equation on the moments $m_n:P_n\rightarrow\mathbb R$. In fact, the very existence of the hierarchy of LPs for the stochastic Ising model is due to locality, where invariance equations involve only the nearest neighbor expressions. It is also worth mentioning that we could replace the invariance condition by reversibility condition:
\ie\label{BS1rev}
m_n\left(d^f_i\right)=0,~~\forall i\in D_{n-1},~\forall f\in P_n.
\fe
Theorem \ref{invGibbs} implies that this condition is obeyed by any invariant measure of the stochastic Ising model with the transition rate $c(i,s)$ respecting the symmetries of the lattice. We will see later that the invariance condition is already sufficient for the convergence of $BS_1(p)$ and the resulting measure will be not only invariant, but also reversible (which is equivalent to Gibbs).

Secondly, the above LP problem $LP(p,n)$ is always feasible because the measure $g_f$ in (\ref{gfinite}) for the statistical Ising model on a large enough but finite torus will satisfy all the conditions. Of course, the Gibbs measure of the statistical Ising model on the infinite lattice (whose existence was established long time ago) also satisfies all the conditions of $LP(p,n)$ for any $n$.

Let us compare $BS_1(p)$ to the traditional ${\cal K}$-moment problem \cite{Schmüdgen1991}, where there will be a variable $x_i\in\mathbb R$ at each lattice site and the moment $m'$ will map polynomials in $x_i$ (of any positive integer power) to $\mathbb R$. $x_i^2=1$ will then be imposed by $m'\left(\left(\left(x_i\right)^2-1\right) f(x)\right)=0$ for all $i\in\Lambda$ and all sums of squares functions $f(x)$. This is indeed how SDP was formulated for 0-1 problem in \cite{10.1007/3-540-45535-3_23} for example. $BS_1(p)$ instead imposes $x_i^2=1$ directly within $m'(\cdot)$ and thus considers polynomials which are at most linear in each $x_i$.

$\langle p\rangle_n^*$ for any $n$ provides a rigorous lower bound on the expectation value $\langle p\rangle$ of any invariant measure respecting all the symmetries of the lattice, for the stochastic Ising model with the transition rate $c(i,s)$. One may use any $c(i,s)$ for the stochastic Ising model as long as it allows for a polynomial expression, and still obtain rigorous lower bounds on the expectation value $\langle p\rangle$. Of course, one can obtain rigorous upper bounds simply by studying the analogous LP problem of maximizing $m_n(p)$.

All the conditions of $LP(p,n)$ are a subset of the conditions of $LP(p,k)$ when $k\geq n$. Therefore, the obtained lower bounds can only increase as we increase the level $n$ of the $LP$ hierarchy: $\langle p\rangle_n^*\leq \langle p\rangle_k^*$, $\forall k\geq n$. Later, we will discuss its convergence to the expectation value of an extremal Gibbs measure.

\subsection{Bootstrapping the Gibbs measure of the statistical Ising model}
In this subsection, we review the bootstrap problem $BS_2'$ proposed in \cite{Cho:2022lcj} for the Gibbs measure of the statistical Ising model and discuss the related bootstrap problem $BS_2$ which will be shown to converge later. $BS_2'$ is mainly based on two properties of the Gibbs measure: reflection positivity and spin-flip equations, both of which are explained in full details in \cite{Cho:2022lcj}. We provide a brief summary of the two below.

For the lattice $\Lambda={\mathbb Z}^d=\{\sum_{\mu=1}^d v_\mu e_\mu,~v_\mu\in\mathbb Z\}$ where $e_\mu$ is the unit vector along the $\mu$-th direction, there are three inequivalent reflections preserving the lattice up to rotations and translations by integer units (except for $d=1$ where there are only two inequivalent reflections). They are denoted as $R_{v,c}$ where the pair $(v,c)$ consists of a vector $v$ on the lattice and a constant $c$. Their actions on a site $i\in\Lambda$ are given by $R_{v,c}(i)=i-{2(v\cdot i-c)\over v^2}v.$ Each of reflections splits $\Lambda$ into half-spaces $H_{v,c}=\{i\in\Lambda~|~v\cdot i\geq c\}$. Three inequivalent reflections are given by $R_{v,c}$ with $(v,c)\in\kappa:=\{(e_1,0),(e_1,1/2),(e_1+e_2,0)\}$, where the last reflection is absent for $d=1$. Reflection positivity states that the expectation value $\langle \cdot\rangle$ of the Gibbs measure satisfies:
\ie
\langle {\cal O}{\cal O}^{R_{v,c}}\rangle\geq0,~~\text{where}~~ {\cal O}=\sum_{A\subset H_{v,c}}t^A\underline{s}_A,~{\cal O}^{R_{v,c}}=\sum_{A\subset H_{v,c}}t^A\underline{s}_{R_{v,c}(A)},~\forall t^A\in\mathbb R,~\forall (v,c)\in\kappa.
\fe

Spin-flip equations can be most easily seen from the Gibbs measure on the finite lattice $g_f$ in (\ref{gfinite}). When evaluating the expectation value of a function using $g_f$, sum over all possible spin configurations $\{u_i\}_{i\in\Lambda}$ is performed. Since the spin values $u_i$ at each site $i\in\Lambda$ are summed over both $-1$ and $1$, the expectation value should be the same if one takes a change of variable $u_i\rightarrow -u_i$. This produces spin-flip equations, which can be extended to the infinite lattice case:
\ie
\bigg\langle {\tilde f}(s)-{\tilde f}\left({\bar s}^i\right)\exp{\left(-2hs_i-2J\sum_{j\in n(i)}s_is_j\right)} \bigg\rangle=0,~~\forall {\tilde f}(s)\in P(S),~\forall i\in\Lambda.
\fe
We now define the bootstrap problem $BS_2$, which is a small extension of the bootstrap problem $BS_2'$ in \cite{Cho:2022lcj}, for the Gibbs measure as follows:

\begin{define}\label{defBS2}
Given $p\in P_m$ for some $m\in{\mathbb N}$, we define the bootstrap problem $BS_2(p)$ as the following hierarchy of SDPs:
\\
$~~~$ For each $n\in{\mathbb N}$ (called the level of the $SDP$ hierarchy) such that $n\geq m$, we have the SDP problem $SDP(p,n)$ of minimizing $m_n(p)$ over the space of candidate moments $m_n:P_n\rightarrow{\mathbb R}$ satisfying the following conditions:
\\
\textbullet~ \textbf{Reflection positivity.} For each of reflections $R_{v,c}$ with $(v,c)\in\kappa\cup\{(e_1+e_2,1)\}$, define the matrix ${\cal M}_n^{v,c}$ by its matrix elements $\left({\cal M}_n^{v,c}\right)_{A,B}=m_n\left(\underline{s}_{(A\cup R_{v,c}(B))\setminus(A\cap R_{v,c}(B))}\right)$ where $A\subset(D_n\cap H_{v,c})$ and $B\subset (D_n\cap H_{v,c})$. Then these matrices should satisfy reflection positivity ${\cal M}_n^{v,c}\succeq0$.
\\
\textbullet~ \textbf{Probability bound.} For all the spin assignments $\{u_i\}_{i\in D_n}$ over $D_n$, $0\leq m_n\left(F\left(\{u_i\}_{i\in D_n},s \right)\right)$ where $F\left(\{u_i\}_{i\in D_n},s \right)$ is the corresponding indicator function.
\\
\textbullet~ \textbf{Linearity.} Given any polynomials $q_1\in P_n$ and $q_2\in P_n$, with $\lambda\in\mathbb R$, their moments satisfy linearity: $m_n(q_1+\lambda q_2)=m_n(q_1)+\lambda m_n(q_2)$.
\\
\textbullet~ \textbf{Unit normalization.} $m_n(1)=1.$
\\
\textbullet~ \textbf{Symmetry.} For any $A\subset D_n$ and $B\subset D_n$ such that $A\sim B$, $m_n(\underline{s}_A)=m_n(\underline{s}_B)$.
\\
\textbullet~ \textbf{Spin-flip equation.} For all $i\in D_{n-1}$ and ${\tilde f}\in P_{n}$, the moments satisfy spin-flip equations:
\ie\label{spinflieEqn}
m_n\left({\tilde f}(s)\right)=m_n\left({\tilde f}\left({\bar s}^i\right)\exp{\left(-2hs_i-2J\sum_{j\in n(i)}s_is_j\right)} \right),
\fe
where the RHS is a polynomial moment due to $s_j^2=1,~\forall j\in\Lambda$.
\\
\\
$~~~$ The minimum of $m_n(p)$ obtained by $SDP(p,n)$ will be denoted as $\langle p\rangle_n^\#$. The corresponding candidate moments $m_n(q)$ for polynomials $q\in P_n$ realizing such a minimum (which may not be unique) will be denoted as $\langle q\rangle_n^\#$.

\end{define}

Bootstrap problem $BS_2'$ in \cite{Cho:2022lcj} is the same as $BS_2$ except that the condition of probability bound was not imposed. It can be checked that reflection positivity alone does not imply probability bound within the domain $D_n$. In $d=1$, the combination of reflection positivity and spin-flip equations still does not imply probability bound. In contrast in $d=2$, it was empirically observed in \cite{Cho:2022lcj} that the same combination implies square positivity which we will later show to be equivalent to probability bound. In any case, adding probability bounds to the SDP does not increase the computational cost significantly since they are merely a lot of $1\times1$ inequalities, rather than a large irreducible matrix inequality.

Similar to the previous discussion on $BS_1$, the existence of the SDP hierarchy for $BS_2$ is due to the local nature of spin-flip equations which involve only the nearest neighbor expressions. Also, the feasibility of $BS_2$ is guaranteed due to the existence of the Gibbs measure on the infinite lattice. The sequence of the mimina $\langle p\rangle_n^\#$ gives rigorous lower bounds which can only increase as $n$ increases. In \cite{Cho:2022lcj}, it was observed that well away from the criticality in $d=2$, $BS_2'$ produces lower and upper bounds for the nearest spin correlator $\langle s_is_{i+e_1}\rangle$ which are very close to each other already at $n=2$, where the gap between the two sometimes was as small as $10^{-15}$.

$BS_1$ and $BS_2$ differ in terms of the equations imposed on the candidate measure, and the latter further imposes reflection positivity. Nonetheless, they should be compatible because the Gibbs measure on the infinite lattice provides a feasible solution to both of them. By Theorem \ref{reverGibbs}, one may expect that $BS_2$ is stronger than $BS_1$ since every Gibbs/reversible measure is invariant.

\begin{lemma}\label{spinflipInv}
For each $n\in\mathbb N$, spin-flip equations of $SDP(p,n)$ include reversibility equations (\ref{BS1rev}), which also include invariance equations of $LP(p,n)$, under the linearity assumption.
\end{lemma}
\textit{Proof}) Making the following choice of $\tilde f$ in spin-flip equations (\ref{spinflieEqn}),
\ie
{\tilde f}(s)=c(i,s)f\left({\bar s}^i\right)
\fe
for $f(s)\in P_n$ and $i\in D_{n-1}$, and using that $c(i,s)\exp{\left(hs_i+J\sum_{j\in n(i)}s_is_j\right)}$ is even in $s_i$ by definition, it is straightforward to derive
\ie
m_n\left( c(i,s)\left(f({\bar s}^i)-f(s)\right) \right) = 0,
\fe
which is the reversibility equation (\ref{BS1rev}). The latter then implies invariance equations by linearity. ~~~~$\blacksquare$

Lemma \ref{spinflipInv} shows that any solution of $SDP(p,n)$ is feasible for $LP(p,n)$. In particular, $\langle p\rangle_n^*\leq\langle p\rangle_n^\#$. It should be noted though that there are considerably many more spin-flip equations than invariance equations at each level of the hierarchy, which may lead to a bigger scale separation issue for $BS_2$ (this issue will be discussed further in section \ref{sec:practice}). In contrast, even if $LP(p,n)$ is further equipped with spin-flip equations, it is safer from the scale separation issue since LP is less sensitive about it than SDP in general. We will consider different combinations of positivity and equations later in section \ref{sec:practice}.

\section{Asymptotic convergence of $BS_1$}\label{sec:convergence}
In this section, we show that as the level $n$ of the LP hierarchy $BS_1$ increases, one can find a convergent subsequence of moments $\{\langle q \rangle_n^*\}_{n\in\mathbb N}$ for $q\in P(S)$ where the convergent limit corresponds to the moments of an invariant measure of the stochastic Ising model. Theorem \ref{invGibbs} then implies that this measure is also a Gibbs measure. Also, $BS_2$ converges in the same sense by Lemma \ref{spinflipInv}.

There are two steps in the proof. The first step is to show that the candidate moments indeed come from a valid probability measure, a problem often called "the moment problem." The second step is to make sure that such a measure is indeed an invariant measure of the stochastic Ising model respecting the symmetries of the lattice. We will obtain the desired result by explicitly constructing a probability measure realizing the candidate moments produced by LP. Since the indicator functions corresponding to the generators of the event space $V$ are finite polynomials, the value of the measure evaluated on such events can be naturally associated with the candidate polynomial moments of the corresponding indicator functions obtained from LP. This natural prescription indeed will be shown to define a consistent probability measure.

\subsection{Moment problem on $S$}\label{subsec:momentProblem}
Establishing a moment problem over a general sample and event space is very difficult and the answers are known only in some special cases, such as Hamburger moment problem or ${\cal K}$-moment problem. In this subsection, we will see that statistical mechanical systems are particularly well-suited for formulating the moment problem.\footnote{Discussions on the moment problem of the statistical mechanical systems can be found for example in \cite{2005math.....12191L}.} Even though we present only the case of the Ising model, the ideas can be straightforwardly generalized to other statistical mechanical systems.

We begin by explaining the moment problem on a finite lattice.\footnote{An equivalent problem was discussed in \cite{2005math.....12191L}, and similar problems where the sample space is given by a finite product of a finite set appeared in various places, such as 0-1 problem and MAX-CUT problem - see e.g. \cite{10.1007/3-540-45535-3_23}.}

\begin{theo}\label{momentsFinite}
Consider a finite subset $\Lambda_f\subset\Lambda={\mathbb Z}^d$. Denote the space of spin configurations over $\Lambda_f$ by $S_f=\{-1,1\}^{\Lambda_f}$ and the corresponding event space by $V_f$. Let $P(S_f)$ be the space of polynomials of spin variables over $S_f$. A candidate moment $m_{\Lambda_f}:P(S_f)\rightarrow\mathbb R$ is a moment of a probability measure over the sample space $S_f$ and the event space $V_f$ if and only if it satisfies:
\\
\textbullet~ \textbf{Probability bound.} For all the spin assignments $\{u_i\}_{i\in \Lambda_f}$ over $\Lambda_f$, $0\leq m_{\Lambda_f}\left(F\left(\{u_i\}_{i\in \Lambda_f},s \right)\right)$ where $F\left(\{u_i\}_{i\in \Lambda_f},s \right)$ is the corresponding indicator function.
\\
\textbullet~ \textbf{Linearity.} Given any polynomials $q_1\in P(S_f)$ and $q_2\in P(S_f)$, with $\lambda\in\mathbb R$, their moments satisfy linearity: $m_{\Lambda_f}(q_1+\lambda q_2)=m_{\Lambda_f}(q_1)+\lambda m_{\Lambda_f}(q_2)$.
\\
\textbullet~ \textbf{Unit normalization.} $m_{\Lambda_f}(1)=1.$

\end{theo}

\textit{Proof}) `` Only if " part is trivial since moments of non-negative functions for a probability measure are non-negative. For the `` if " part, we explicitly construct a probability measure $\R_{\Lambda_f}$ giving rise to the moment $m_{\Lambda_f}$. Since the moment is defined on all polynomials of spin variables over $\Lambda_f$, it is defined in particular on the indicator functions (\ref{indicator}): $F\left(\{u_i\}_{i\in A},s \right)=\prod_{i\in A}\left({1+u_is_i\over2}\right)$ for all $A\subset\Lambda_f$. The event space $V_f$ is generated by the events $E\left(\{u_i\}_{i\in A},s \right)$ defined in (\ref{eventDef}), where spin assignments $u_i$ are specified over a subset $A\subset\Lambda_f$. We define $\R_{\Lambda_f}$ by its value on these generating events:
\ie
\R_{\Lambda_f}\left(E\left(\{u_i\}_{i\in A} \right)\right):=m_{\Lambda_f}\left(F\left(\{u_i\}_{i\in A},s \right)\right)=m_{\Lambda_f}\left(\prod_{i\in A}\left({1+u_is_i\over2}\right)\right).
\fe
We extend the definition linearly: given disjoint generating events $\bigg\{E\left(\{u^{(t)}_i\}_{i\in A^{(t)}} \right)\bigg\}_{t\in T}$ for some finite index set $T$ such that $A^{(t)}\subset\Lambda_f~~\forall t\in T$,
\ie
\R_{\Lambda_f}\left(\bigcup_{t\in T}E\left(\{u^{(t)}_i\}_{i\in A^{(t)}} \right)\right)=\sum_{t\in T}\R_{\Lambda_f}\left(E\left(\{u^{(t)}_i\}_{i\in A^{(t)}} \right)\right).
\fe
$\R_{\Lambda_f}$ on the complement events are defined by
\ie
\R_{\Lambda_f}\left(\left(\bigcup_{t\in T}E\left(\{u^{(t)}_i\}_{i\in A^{(t)}} \right)\right)^c\right)=1-\sum_{t\in T}\R_{\Lambda_f}\left(E\left(\{u^{(t)}_i\}_{i\in A^{(t)}} \right)\right).
\fe
This definition is consistent in that, if there are two sets of pairwise disjoint events such that their unions coincide, $\R_{\Lambda_f}$ evaluated on them are the same. This is due to the assumption $m_{\Lambda_f}(1)=1$, linearity, and the properties of the indicator functions, together with the fact that the sample and event spaces under consideration are finite. This determines $\R_{\Lambda_f}$ completely and finite additivity of $\R_{\Lambda_f}$ naturally follows.

It remains to show that $\R_{\Lambda_f}$ is non-negative and bounded from above by 1. By definition, if we sum over all the indicator functions corresponding to all the events where every spin over $\Lambda_f$ is specified, we should get the function 1:
\ie
\sum_{u\in\{-1,1\}^{\Lambda_f}}F\left(\{u_i\}_{i\in \Lambda_f},s \right)=1.
\fe
Since every summand in the above is non-negative by probability bound assumption, linearity and unit normalization imply that any partial sum of $m\left(F\left(\{u_i\}_{i\in \Lambda_f},s \right)\right)$ should be bounded from above by 1, leading to
\ie
0\leq \R_{\Lambda_f}\left(\bigcup_{t\in T}E\left(\{u^{(t)}_i\}_{i\in A^{(t)}} \right)\right)\leq 1,
\fe
for all pairwise disjoint events $E\left(\{u^{(t)}_i\}_{i\in A^{(t)}} \right)$. Since $m_{\Lambda_f}(1)=1$ by assumption, $\R_f$ evaluated on the complement events are also bounded from below by 0 and from above by 1. This completes the proof. ~~~~$\blacksquare$

Now, we extend probability measure $\R_{\Lambda_f}$ constructed above to a probability measure $\R$ over the sample space $S$ and the event space $V$ on the infinite lattice $\Lambda={\mathbb Z}^d$ using the Kolmogorov extension theorem in stochastic process. The key idea of the extension theorem is that if probability measures defined on the finite subsets of an infinite set are compatible with each other in the sense explained below, then it is guaranteed that there exists a probability measure on the infinite set which agrees with probability measures on the finite subsets when restricted to those finite subsets.

\begin{theo}\label{momentS}
A candidate moment $m:P(S)\rightarrow\mathbb R$ is a moment of a probability measure if it satisfies:
\\
\textbullet~ \textbf{Probability bound.} $0\leq m\left(F\left(\{u_i\}_{i\in A},s \right)\right)$ for any the spin assignments $\{u_i\}_{i\in A}$ over any finite subset $A\subset\Lambda$.
\\
\textbullet~ \textbf{Linearity.} Given any polynomials $q_1\in P(S)$ and $q_2\in P(S)$, with $\lambda\in\mathbb R$, their moments satisfy linearity: $m(q_1+\lambda q_2)=m(q_1)+\lambda m(q_2)$.
\\
\textbullet~ \textbf{Unit normalization.} $m(1)=1.$
\end{theo}

\textit{Proof}) Since $\Lambda={\mathbb Z}^d$ is countable, we can consider the sequence $\{r_i\}_{i\in \mathbb N}$ where $r_i\in\Lambda$ and $r_i\neq r_j$ for $i\neq j$ such that $\bigcup_{i\in \mathbb N}\{r_i\}=\Lambda$. Given $N\in\mathbb N$, consider the subsequence $R_N:=\{r_1,r_2,...,r_N\}$. Considering $R_N$ as $\Lambda_f$ in Theorem \ref{momentsFinite}, we obtain a valid probability measure $\R_{R_N}^{r_1,...,r_N}$ over the sample space $\{-1,1\}^{R_N}$ and the corresponding event space $V_{R_N}$ as defined in the proof of Theorem \ref{momentsFinite}:
\ie
\R_{R_N}^{r_1,...,r_N}\left(E\left(\{u_i\}_{i\in A} \right)\right):=m\left(F\left(\{u_i\}_{i\in A},s \right)\right)=m\left(\prod_{i\in A}\left({1+u_is_i\over2}\right)\right),
\fe
for all $A\subset R_N$ and spin configurations $\{u_i\}_{i\in A}$ over it. This definition extends linearly and specifies the probability measure $\R_{R_N}^{r_1,...,r_N}$ completely as outlined in the proof of Theorem \ref{momentsFinite}. Given any permutation $\pi$ on the set $\{1,2,...,N\}$, we similarly define
\ie
\R_{R_N}^{r_{\pi(1)},...,r_{\pi(N)}}\left(E\left(\{u_i\}_{i\in A} \right)\right):=m\left(\prod_{i\in A}\left({1+u_{\pi(i)}s_{\pi(i)}\over2}\right)\right),
\fe
and this defines a valid probability measure $\R_{R_N}^{r_{\pi(1)},...,r_{\pi(N)}}$. These probability measures are then manifestedly permutation invariant.

Furthermore, given any $N'>N$,
\ie
\R_{R_{N'}}^{r_1,...,r_{N'}}\left(E\left(\{u_i\}_{i\in A} \right)\right)=m\left(F\left(\{u_i\}_{i\in A},s \right)\right)=\R_{R_N}^{r_1,...,r_N}\left(E\left(\{u_i\}_{i\in A} \right)\right)
\fe
for any $A\subset R_N$ and spin configurations $\{u_i\}_{i\in A}$ over it. This implies that given the joint probability measure $\R_{R_N}^{r_1,...,r_{N'}}$, the marginal probability measure where the spin values on $\{r_{N+1},...,r_{N'}\}$ are summed over is given by $\R_{R_N}^{r_1,...,r_N}$.

The above two properties of $\R_{R_N}^{r_1,...,r_N}$, permutation invariance and marginality, are the sufficient conditions for the Kolmogorov extension theorem, which states that there is a probability measure $\R$ over the sample space $S$ and the event space $V$ on the infinite lattice $\bigcup_{i\in \mathbb N}\{r_i\}=\Lambda={\mathbb Z}^d$ such that its marginals are given by $\R_{R_N}^{r_1,...,r_N}$:
\ie
\R\left(E\left(\{u_i\}_{i\in A} \right)\right)=m\left(F\left(\{u_i\}_{i\in A},s \right)\right)=\R_{R_N}^{r_1,...,r_N}\left(E\left(\{u_i\}_{i\in A} \right)\right),
\fe
for all $N\in\mathbb N$, $A\subset R_N$, and spin configurations $\{u_i\}_{i\in A}$ over $A$. By construction, $m:P(S)\rightarrow\mathbb R$ is the moment of the probability measure $\R$. ~~~~$\blacksquare$

Probability bounds are the minimal positivity requirements for the existence of a measure realizing the candidate moments. It turns out that they are equivalent to another familiar positivity condition, square positivity.

\begin{lemma}\label{lem:probSqr}
Given a candidate moment $m_{\Lambda_f}:P(S_f)\rightarrow\mathbb R$ satisfying the linearity and unit normalization of Theorem \ref{momentsFinite}, the following two conditions are equivalent:\\
\textbullet~ \textbf{Probability bound.} For all the spin assignments $\{u_i\}_{i\in \Lambda_f}$ over $\Lambda_f$, $0\leq m_{\Lambda_f}\left(F\left(\{u_i\}_{i\in \Lambda_f},s \right)\right)$ where $F\left(\{u_i\}_{i\in \Lambda_f},s \right)$ is the corresponding indicator function.
\\
\textbullet~ \textbf{Square positivity.} For any polynomial $q\in P(S_f)$, the moment of its square is positive: $m_{\Lambda_f}(q^2)\geq0$.
\end{lemma}

\textit{Proof}) Since every indicator function squares to itself, square positivity trivially implies probability bound. For the opposite direction, note that $F\left(\{u_i\}_{i\in \Lambda_f},s \right)$ for all $\{u_i\}_{i\in \Lambda_f}$ provide a complete basis of $P(S_f)$. Therefore, we can expand
\ie
q=\sum\limits_{u\in\{-1,1\}^{\Lambda_f}}t_uF\left(\{u_i\}_{i\in \Lambda_f},s \right)
\fe
for any $q\in P(S_f)$ with $t_u\in\mathbb R$. By definition, the product of indicator functions corresponding to pairwise disjoint events vanishes. Therefore,
\ie
m_{\Lambda_f}\left(q^2\right)=\sum\limits_{u\in\{-1,1\}^{\Lambda_f}}\left(t_u\right)^2m_{\Lambda_f}\left(F\left(\{u_i\}_{i\in \Lambda_f},s \right)\right)\geq0,
\fe
which is the desired result. ~~~~$\blacksquare$

Similarly, Theorem \ref{momentS} holds true if probability bound is replaced by square positivity. Lemma \ref{lem:probSqr} implies that the LP problem $BS_1$ can be equivalently formulated as a SDP problem where square positivity is imposed instead of probability bound. This is because, by defining the matrix ${\cal M}$ via its matrix elements ${\cal M}_{A,B}=m\left(\underline{s}_{(A\cup B)\setminus(A\cap B)}\right)$ where $A\subset \Lambda_f$ and $B\subset \Lambda_f$, square positivity is equivalent to ${\cal M}\succeq0$, which is a SDP constraint. However, there is no advantage in doing so because LP is much faster and cheaper than the equivalent SDP in this case.

Theorem \ref{momentS} not only shows the existence of a probability measure $\R$ realizing the candidate moments, but also is constructive in that $\R$ evaluated on any event can be expressed in terms of the moments of the indicator functions. For example, given an infinite sequence of disjoint events $\{E_k\}_{k\in\mathbb N}$, $\R$ evaluated on the partial union $\bigcup_{k=1}^n E_n$ is bounded from above by 1. Therefore, the limit $\R\left(\bigcup_{k=1}^n E_n\right)$ as $n\rightarrow\infty$ exists and is what the countable additivity of $\R$ predicts. Similarly, given an infinite sequence of strictly descending events $E_1\supsetneq E_2\supsetneq E_3 \supsetneq...$, the sequence $\R(E_k)$ is non-increasing and bounded from below by 0. Therefore, the limit $\R(E_k)$ as $k\rightarrow\infty$ exists and this for example defines the value of $\R$ evaluated on the event where spin values on an infinitely many lattice sites are specified. Even though we expect such a value to be essentially 0 for the Ising model, it may even be 1 for extreme cases like Dirac measure on $S$. This illustrates the point that the moment problem we discussed above is about the space of all possible probability measures on the sample space $S$ and the event space $V$, while the probability measure of our interest is specifically that of the Ising model. We now address how the symmetry and invariance conditions of $BS_1$ pin down the invariant/reversible/Gibbs measure of the Ising model within the space of all probability measures on $S$.

\subsection{Asymptotic convergence of the Ising bootstrap}
In this subsection, we show that the bootstrap problem $BS_1$ converges as the level $n$ of the LP hierarchy increases. Two main ingredients for the proof have already been presented: the moment problem in Theorem \ref{momentS} and the polynomial representation of invariance equations in Lemma \ref{invEqPoly}. The rest of the proof follows the usual steps.\footnote{Similar proofs of convergence of SDP hierarchy for the dynamical system or Markov chain bootstrap can be found for example in \cite{hernández2012markov,2018arXiv180708956K}. In \cite{Kazakov:2021lel}, a similar proof of convergence of the bootstrap method for the quartic one-matrix model was presented based on the result of the Hamburger moment problem. We thank the anonymous referee of the Journal of High Energy Physics for pointing this reference to us.}

\begin{theo}\label{BS1convergence}
Consider the bootstrap problem $BS_1(p)$ with LP hierarchy $LP(p,n)$ for $p\in P_m$. Recall that the minimum of $m_n(p)$ obtained by $LP(p,n)$ is denoted as $\langle p\rangle_n^*$ and the corresponding candidate moments $m_n(q)$ of the polynomials $q\in P_n$ are denoted as $\langle q\rangle_n^*$. For $l\in\mathbb N$, define the sequence $\mathbb{N}_{m,l}=\{\max(m,l),\max(m,l)+1,\max(m,l)+2,...\}$. Then,
\\
\textbullet~ The limit $\langle p\rangle_\infty:=\lim\limits_{n\rightarrow\infty}\langle p\rangle_n^*$ exists.
\\
\textbullet~ Given $l\in\mathbb N$, consider the polynomials $\underline{s}_A\in P_l$ for $A\subset D_l$. The sequence $\{\langle \underline{s}_A\rangle_n^*\}_{n\in\mathbb{N}_{m,l}}$ in $\mathbb{R}^{2^{|D_l|}}$ has a convergent subsequence $\{\langle \underline{s}_A\rangle_n^*\}_{n\in Q}$ for an appropriate index set $Q\subset\mathbb{N}_{m,l}$, whose convergent limit is denoted as $\langle \underline{s}_A\rangle_\infty:=\lim\limits_{n\in Q,n\rightarrow \infty}\langle \underline{s}_A\rangle_n^*$.

Furthermore, there exists an invariant measure $\R$ of the stochastic Ising model with the transition rate $c(i,s)$ which respects the lattice symmetries, whose corresponding expectation values $\langle\cdot\rangle$ satisfy:
\\
\textbullet~ $\langle p\rangle = \langle p\rangle_\infty$.
\\
\textbullet~ $\langle \underline{s}_A\rangle = \langle \underline{s}_A\rangle_\infty$.

Finally, given any other invariant measure $\R'$ of the stochastic Ising model with the transition rate $c(i,s)$ respecting the lattice symmetries, $\langle p \rangle \leq \langle p \rangle'$, where $\langle p \rangle'$ is the expectation value of $p$ given by $\R'$.
\end{theo}

\textit{Proof}) Square positivity (which follows from probability bounds by Lemma \ref{lem:probSqr}) and unit normalization imply that $-1\leq\langle\underline{s}_A\rangle_n^*\leq1$ for any $A\subset D_l$ and any $n\in\mathbb{N}_{m,l}$. Therefore, $\{\langle \underline{s}_A\rangle_n^*\}_{n\in\mathbb{N}_{m,l}}$ is a bounded sequence in $\mathbb{R}^{2^{|D_l|}}$ and thus has a convergent subsequence $\{\langle \underline{s}_A\rangle_n^*\}_{n\in Q}$ with the limiting values $\langle\underline{s}_A\rangle_\infty$. By continuity, $\langle\underline{s}_A\rangle_\infty$ as candidate moments satisfy all the conditions of Theorem \ref{momentsFinite} with $\Lambda_f=D_l$. Therefore, we can construct a probability measure $\R_l$ on the sample space $\{-1,1\}^{D_l}$ and corresponding event space by declaring that its moments are given by $\langle\underline{s}_A\rangle_\infty$, $A\subset D_l$. Furthermore, for all $l_2>l_1\geq l$, we can similarly define $\R_{l_1}$ and $\R_{l_2}$ such that $\R_{l_1}$ is the marginal probability measure of $\R_{l_2}$. Then, following the proof of Theorem \ref{momentS}, there is a probability measure $\R$ on the sample space $S$ and the event space $V$ such that its marginal probability measures are \{$\R_{l'}\}_{l'\geq l}$. Since each lattice symmetry constraint involves only finitely many moments, $\R$ respects the lattice symmetries by continuity. Similarly, each invariance equation involves only finitely many moments and thus the moments of $\R$ satisfy invariance equations in Lemma \ref{invEqPoly} with the transition rate $c(i,s)$ by continuity. Therefore, $\R$ is the invariant measure of the stochastic Ising model with the transition rate $c(i,s)$ and the corresponding expectation value $\langle \underline{s}_A\rangle$ of $\underline{s}_A$ for any $A\subset D_l$ agrees with that given by the finite marginal probability measure: $\langle \underline{s}_A\rangle = \langle \underline{s}_A\rangle_\infty$.

As discussed below Definition \ref{defBS1} of $BS_1(p)$, the sequence $\{\langle p\rangle_n^*\}$
is a non-decreasing sequence in $\mathbb R$. Square positivity, unit normalization, and linearity also imply that the sequence is bounded from above. Therefore, its limit $\langle p\rangle_\infty$ exists and coincides with the corresponding moment of $\R$: $\langle p\rangle = \langle p\rangle_\infty$. Let $\nu$ be an invariant measure of the stochastic Ising model with the transition rate $c(i,s)$ respecting the lattice symmetries such that its moment $\langle p\rangle_\nu$ for $p$ is minimal among all such invariant measures. Since $\langle\cdot\rangle_\nu$ is feasible for $BS_1(p)$, we have $\langle p\rangle\leq\langle p\rangle_\nu$. Because $\R$ itself is an invariant measure, the definition of $\nu$ implies $\langle p\rangle \geq\langle p\rangle_\nu$. Therefore, $\langle p\rangle = \langle p\rangle_\nu$. ~~~~$\blacksquare$

A few corollaries follow from previous discussions. Due to the symmetry conditions of $BS_1(p)$, Theorem \ref{invGibbs} implies:
\begin{corollary}
Probability measure $\R$ in Theorem \ref{BS1convergence} is a Gibbs measure of the statistical Ising model.
\end{corollary}

\begin{corollary}
The bootstrap problem $BS_2(p)$ converges in the same sense as $BS_1(p)$ in Theorem \ref{BS1convergence}.
\end{corollary}

It is worth mentioning how to obtain the extremal Gibbs measure from $BS_1(p)$ or $BS_2(p)$. A Gibbs measure is extremal if it cannot be written as a weighted sum of two different Gibbs measures. Of course, this notion is nontrivial only in the low temperature and in the absence of the external magnetic field where there are infinitely many Gibbs measures. This is exactly where the order parameter $\langle s_i\rangle$ (also called the magnetization) becomes nonzero. Therefore, if we choose $p=s_i$ and minimize (or maximize) $\langle s_i\rangle$, the corresponding measure $\R$ is expected to be an extremal measure.

Theorem \ref{BS1convergence} for $BS_1(p)$ may sound strange from the Euclidean field theory perspective since the natural positivity of the latter is reflection positivity, while $BS_1(p)$ converges with just probability bound/square positivity. However, reflection positivity is a property satisfied by specific Hamiltonians and thus, it indirectly appears through invariance equations. Rather surprising fact is that it is spin-flip equations which are analogous to the equations of motions of the Euclidean field theories, while $BS_1(p)$ contains only a "summed" version of such equations of motions. This is another place where the nontriviality of Theorem \ref{invGibbs} is highlighted. Another very curious fact is that, in $d=1$, the combination of reflection positivity and spin-flip equations was not enough to produce probability bounds, while it seems enough for $d=2$ from empirical evidences.

\section{The statistical Ising bootstrap in practice}\label{sec:practice}
In the previous section, we have shown that the bootstrap problems $BS_1$ and $BS_2$ converge in principle. In this section, we discuss how the insights from the convergence proof may help formulating other convergent bootstrap problems and hopefully produce better bootstrap bounds on the expectation values.

\subsection{Improving the LP and SDP}
If we replace invariance equations with spin-flip equations in $BS_1$, we not only obtain stronger bounds (which are still rigorous even for the invariant measures), but also can reduce the number of probability bounds that we need to impose. This is essentially because the transition rate $c(i,s)$ is strictly positive.

\begin{lemma}\label{lem:lessProbBds}
If invariance equations of $BS_1$ are replaced by spin-flip equations, we can reduce probability bound conditions to the following subset and the resulting bootstrap problem still converges:\\
\ie
0\leq m_n\left(F\left(\{u_i\}_{i\in D_n},s\right)\right),
\fe
for all spin assignments $\{u_i\}_{i\in D_n}$ such that $u_i=1$ for $i\in D_{n-1}$.
\end{lemma}
\textit{Proof}) Spin flip equations imply reversibility conditions
\ie
m_n\left(c(i,s)\left(f({\bar s}^i)-f(s)\right)\right) = 0,
\fe
for all $f(s)\in P_n$ and $i\in D_{n-1}$. Taking $f(s)$ to be a specific indicator function $F\left(\{u_j\}_{j\in D_n},s\right)$, reversibility condition becomes
\ie\label{spinflipEqInd}
c(i,u')m_n\left(F\left(\{u'_j\}_{j\in D_n},s\right)\right) = c(i,u)m_n\left(F\left(\{u_j\}_{j\in D_n},s\right)\right),
\fe
where $u'_j=u_j$ for $j\neq i$ and $u'_i=-u_i$. Probability bound $0\leq F\left(\{u_j\}_{j\in D_n},s\right)$ then implies
\ie
0\leq m_n\left(F\left(\{u'_j\}_{j\in D_n},s\right)\right),
\fe
since $c(i,u)$ is strictly positive. By repeatedly applying the same argument, we obtain
\ie
0\leq m_n\left(F\left(\{u''_j\}_{j\in D_n},s\right)\right),
\fe
for all $u''$ such that $u''_j=u_j$ for $j\in\partial D_n$.~~~~$\blacksquare$

This Lemma implies that the number of probability bounds which should be imposed is of order $2^{|\partial D_n|}\sim 2^n$ rather than $2^{|D_n|}\sim 2^{n^2}$ in the presence of reversibility conditions. Instead, the number of spin-flip equations is of order $|D_{n-1}|2^{|D_n|}\sim n^2 2^{n^2}$ while that of invariance equations is of order $|D_{n-1}|\sim n^2$. Therefore, the size of the LP increases from $2^{n^2}$ to $n^22^{n^2}$ as we replace invariance equations with spin-flip equations, but this replacement nonetheless produces stronger bounds.

Lemma \ref{lem:lessProbBds} also applies to the SDP problem of $BS_2$. There is even a further reduction in the number of probability bounds since reflection positivity implies that the indicator function corresponding to reflection symmetric spin assignments has a non-negative moment. Therefore, one only needs to impose probability bounds on the spin assignments over $\partial D_n$ which are not symmetric under all of the reflections.

\subsection{Comparisons of different bootstrap approaches}
We have discussed two sets of positivities in this work for the LP/SDP hierarchy (for each domain $D_n\subset\Lambda$):
\begin{gather*}
\text{Probability bound (LP)}~\subset~\text{Reflection positivity + Probability bound (SDP)}
\end{gather*}
These positivities are sufficient to solve the moment problem on $S$. We then combine one of these with the equations specifying the statistical/stochastic Ising model:
\begin{gather*}
\text{Invariance equations}~\subset~\text{Spin-flip equations}.
\end{gather*}

Any combination of positivity and equations in the above is guaranteed to converge. LP is much faster and cheaper than SDP, but the latter involving reflection positivities produces stronger bounds. Including too many equations leads to a SDP matrix whose ratio between the element of the biggest magnitude to the element of the smallest nonzero magnitude is large. In such cases, higher precision SDP solvers are needed which are necessarily much slower. Therefore, there is an advantage in using invariance equations instead of spin-flip equations because such a scale problem may be milder for the former. For the LP problem in contrast, such a precision issue is less likely to occur and imposing more equations do not require much extra computation cost. One great advantage of LP is that equations do not need to be solved because they can be directly implemented as part of the linear constraints. In contrast, directly incorporating equations into SDP is hard in practice, and one should instead solve the equations and substitute the solutions into SDP matrices by hand.

In \cite{Cho:2022lcj}, it was observed that $BS_2'$ produces the weakest bounds around the critical points. We thus take the $d=2$ Ising model at the criticality, $J={\log(1+\sqrt{2})\over2},~h=0$, as the testing ground for different combinations of positivities and equations, where the objective function was the free energy $\langle p\rangle=\langle s_is_{i+e_1}\rangle$ whose exact value is given by $0.707107...$. The following table provides a summary of the results obtained by MOSEK \cite{mosek} on the Intel i9-10900F processor. The abbreviations are given by:
\\
$~~~~~$ P: positivity, E: equations, n: LP/SDP hierarchy level, Min: lower bound on $\langle p\rangle$ rounded down to six significant digits, Max: upper bound on $\langle p\rangle$ rounded up to six significant digits, ST: solver runtime, PB: probability bound, RP: reflection positivity, I: invariance equations based on the transition rate $c^*(i,s)$, S: spin-flip equations.

\begin{center}
\begin{tabular}{||c c c c c c c||} 
 \hline
 P & E & n & Min & Max & ST & Note\\ [0.5ex] 
 \hline\hline
 PB & I & 3 & 0.167853 & 0.851084 & $\sim$0.5 sec & ~\\ 
 \hline
 PB & S & 3 & 0.303045 & 0.820244 & $\sim$0.5 sec & ~\\ 
 \hline
 PB & S & 3.5 & 0.444667 & 0.820244 & a few mins & only a subset of PB used\\ 
 \hline
 RP & I & 3 & 0.628600 & 0.753475 & a few secs & ~\\ 
 \hline
 RP & S & 3 & 0.654752 & 0.753475 & a few mins & data from \cite{Cho:2022lcj}\\ 
 \hline
 RP & I & 4 & 0.682418 & 0.740840 & $\sim$20 mins & only a subset of RP used
 \\ [0.5ex] 
 \hline
\end{tabular}
\end{center}

For the third row, we imposed spin-flip equations for polynomials in $P_3$ where the spin flip may take place at the boundary of $D_3$. PB was then imposed only on the spin configurations generated by such spin-flip equations. For the last row, we truncated reflection positivity matrices to some arbitrary $200\times200$ principal submatrices because the full problem was slow. As expected, LP (used for PB) is much faster than SDP (used for RP), but produces much weaker bounds than the latter. However, it seems straightforward to extend the LP to $D_4$, in which case the bounds may be comparable to those obtained by SDP while still requiring shorter amount of runtime for the solver.

For SDP, spin-flip equations on $D_3$ produced SDP matrices where the element of the biggest magnitude was $\sim10^3$, while it was $\sim10^2$ for invariance equations on $D_3$. Even though there are only 5 invariance equations on $D_3$ (fourth row), they still produce bounds of the same order as the full 549 spin flip equations on $D_3$ (fifth row), where the upper bounds are identical and the solver runtime is much shorter. This is where Theorem \ref{invGibbs} is realized in practice. Finally, invariance equations on $D_4$ were still mild enough in terms of the scaling to produce SDP matrices that can be run on a double-precision solver and produced the strongest bounds (last row).

\section{Discussions}\label{sec:diss}
In this work, we discussed the convergence of the bootstrap approach to the statistical and stochastic Ising model. We discuss several interesting conclusions.

\textbullet~ As already demonstrated many times in literature (e.g. \cite{hernández2012markov,2015arXiv151205599F,2018arXiv180708956K}) and again in this work, Markov processes and stochastic models are amenable to the bootstrap approach. This is essentially because the observable of interest in these systems is an invariant measure, and LP/SDP provide systematic methods to study such a measure problem. One great feature manifest in many of such systems is that they have better chances to be ergodic and free of special solutions. This is in contrast to the classical dynamical systems where chaotic systems are always accompanied by infinitely many unstable periodic orbits which prevent bootstrap from directly accessing the ergodic orbit. Furthermore, this work suggests that any system that used to be studied by the traditional MCMC simulations may allow for an alternative bootstrap approach - one may choose to run the simulations, or to "bound" the simulations. The latter may be more expensive computationally, but the relative advantage is that bootstrap provides rigorous bounds on the observables of the infinite volume systems directly.

\textbullet~ We also demonstrated that statistical mechanical systems on the lattice are particularly well-suited for the Lasserre hierarchy formulation. As long as there is a notion of compactness on the local degrees of freedom and there is locality in the system, most of the steps in the moment problem and the convergence presented in this work may be extended straightforwardly. For example, lattice pure Yang-Mills theory may be an interesting case to study, where the compactness is present since $SU(N)$ is compact.\footnote{Large $N$ pure Yang-Mills theory on the lattice has recently been studied in \cite{Kazakov:2022xuh}. The approach seems to allow for a straightforward generalization to the finite $N$ case. We thank Zechuan Zheng for the relevant discussion.} Above all, the very definition of the Gibbs measure on the infinite lattice using the local conditional probabilities allows for a very natural bootstrap formulation.

\textbullet~ A general lesson for the positive measure bootstrap is that considering the associated MCMC may help identifying the relevant pieces of bootstrap conditions. In the case of the Ising model considered in this work, there are plethora of spin correlator inequalities (some of which are non-convex) which have played important roles in establishing highly nontrivial results such as the existence of the phase transition. Also, the number of spin-flip equations explodes as the domain under consideration increases. Considering the problem of finding the invariant measure of the stochastic Ising model showed that the minimal set of bootstrap conditions which guarantee the convergence is probability bounds and invariance equations. In other words, these are enough to completely determine the theory. Of course for more general theories, the analogue of Theorem \ref{invGibbs} may be hard to prove and the set of invariant measures may be strictly bigger than the set of physical measures of interest. Still, bootstrap approach may provide insights into such differences which are interesting problems on their own.

There are also very obvious next steps.

\textbullet~ It will be very important to obtain the rate of the convergence as $n$ increases. At least away from the criticality, empirical results of \cite{Cho:2022lcj} suggest that the convergence is exponentially fast. Establishing the rate of the convergence is meaningful from both conceptual and practical perspectives. The asymptotic convergence shows that bootstrap can serve as an alternative definition of the system, while the rate of the convergence will tell us how to determine the physical observables to any desired precision. It will be also interesting to understand how much reflection positivity speeds up the convergence.

\textbullet~ In many examples on the lattice, an important quantity which is not explored in this work is the long-range correlators, which are often used to extract critical exponents or mass gap. From the convergence proof of $BS_1$, we learned that to pin down the invariant measure, we need to impose probability bounds and invariance equations over the entire lattice in principle. If we consider a subset of probability bounds and invariance equations involving the long-range correlators, the bounds will be tight only if there is some universality among all the measures satisfying the subset of conditions. Furthermore, we will need to face the computational cost which increases exponentially as the number of spin configurations to be considered grows. Whether there will be an alternative approach to directly study critical exponents or mass gap within the bootstrap framework is unclear at the moment.

\textbullet~ Given the fundamental importance of reflection positivity and the role it played in showing various properties of the Ising model, it would be desirable to establish the precise relation between the positivity of the Gibbs measure and reflection positivity. Even though reflection positivity is a property of specific Hamiltonians, it is curious that it does not imply probability bounds even in the presence of spin-flip equations in $d=1$ statistical Ising model. At least in this case, the nice inner product structure defined by reflection positivity together with the equations of motions is not be enough to deduce that the candidate moments originate from a valid probability measure. The question readily extends to any reflection-symmetric Gibbs measures in other statistical mechanical systems.

\textbullet~ Needless to say, it is worth improving LP/SDP formulation itself. Indicator functions played a central role in showing the convergence in this work. They also provide a complete basis of $P_n$ and make probability bound and spin-flip equations very simple by definition (see for example (\ref{spinflipEqInd})). The only drawback of this basis is that translation invariance is not straightforward to impose. From the perspective of Theorem \ref{invGibbs}, it may seem that translation invariance is essential, but it is also known that the Gibbs measures of the statistical Ising model are translation invariant. Therefore, one would expect to recover translation invariance by imposing spin-flip equations even if translation invariance is not imposed at the level of bootstrap.

\section*{Acknowledgments}
We greatly appreciate Clay Cordova for persistently asking M.C. about the convergence of the Ising bootstrap, which initiated this work. We would like to thank Hamza Fawzi, Weihao Guo and Zechuan Zheng for helpful discussions, and Hamza Fawzi for helpful comments on the preliminary draft. We also thank the anonymous referee of the Journal of High Energy Physics for suggestions on the draft. M.C. is supported by the Sam B. Treiman Fellowship at the Princeton Center for Theoretical Science. X.S.\ was partially supported by the NSF Career award 2046514, and by a fellowship from the Institute for Advanced Study at Princeton during 2022-2023.

\bibliographystyle{JHEP}
\bibliography{glauber}

\end{document}